\documentclass{aa}
\usepackage{graphicx}
\begin{document}
\title{Spectroscopic investigation of unstudied southern
PNe\thanks{Based on observations collected at the European Southern Observatory, La
Silla}}


\author{
M. Emprechtinger\inst{1}\and T. Forveille\inst{2}$^,$\inst{3} \and S.
Kimeswenger\inst{1}}


\institute{ Institut f{\"u}r Astrophysik der Universit{\"a}t Innsbruck,
Technikerst. 25, A-6020 Innsbruck, Austria \and
Observatoire de Grenoble, BP 53X, 38041 Grenoble Cedex, France 3 \and
CFHT, PO Box 1597, Kamuela, HI 96743, USA}

\date{Received 18 February 2004 / accepted 18 May 2004}

\abstract{We present a spectroscopic investigation of  two
hitherto unstudied galactic planetary nebulae  (\mbox{MeWe 1-10}
and \mbox{MeWe 1-11}) and one candidate object (\mbox{MeWe 2-5}).
The candidate object clearly has been identified as a bipolar
hourglass shaped PN. The galactic foreground extinction was
derived and using photoionization models with CLOUDY the two round
objects were classified as highly evolved nebulae.

}
\maketitle


\section{Introduction}
The two PNe, MeWe 1-10, MeWe 1-11 and the candidate object MeWe
2-5  were discovered on \mbox{ESO-R} film copies during a
systematic survey by Melmer \& Weinberger (\cite{MeWe}). They
classified the first two objects on the basis of morphology as PNe
and the last one to be a candidate. They have never been studied
individually. The same classification then was used by Acker et
al. (\cite{ESO}). The only published measurements are radial
velocities for \mbox{MeWe 1-10} and \mbox{MeWe 1-11} in the
context of a survey of PNe to study the galactic population
(Durand et al. \cite{durand};
Beaulieu et al. \cite{silvie}). \\
We present here a spectroscopic investigation and  photoionization models of
these targets.

\noindent Planetary nebulae (PNe) have for a long time been known
as representing an inescapable bottleneck in late stellar
evolution for stars of intermediate masses. Many studies focus
either on a few well--known prototypes or only give a very coarse
overview (see the statistics in Acker \cite{AA}). We contribute to
the sample of individual studies to get better statistics for
larger PNe samples. A number of the most evolved of them were used
as a probe of PNe--interstellar matter (ISM) interaction (Rauch et
al. \cite{elise1}; Kerber et al. \cite{elise2}, \cite{elise3}).
The spectroscopy presented here shows that MeWe 1-10 and 1-11 also
might be good probes for a detailed deeper study of the ISM
interaction, which is beyond the scope of this research note.

\section{Observations and  data reduction}

The data was obtained on July 16$^{\rm th}$ 1998 using the Danish
1.54\,m telescope with the DFOSC spectrograph at ESO La Silla,
Chile. A LORAL 2k x 2k CCD detector and Grism \#4 with a
resolution of 3.1~nm/pixel and a usable range from 4500~nm to
7000~nm were attached. The calibration was done using usual
procedure in MIDAS and the calibration data from the DFOSC manual.
The standard star EG 274 (Hamuy et al. \cite{STd}) was used for
flux calibration. For slit centering purposes a narrow band
[\ion{O}{III}] image was taken before each spectrum.

\begin{figure*}[ht]
\centerline{\resizebox{11.0cm}{!}{\includegraphics{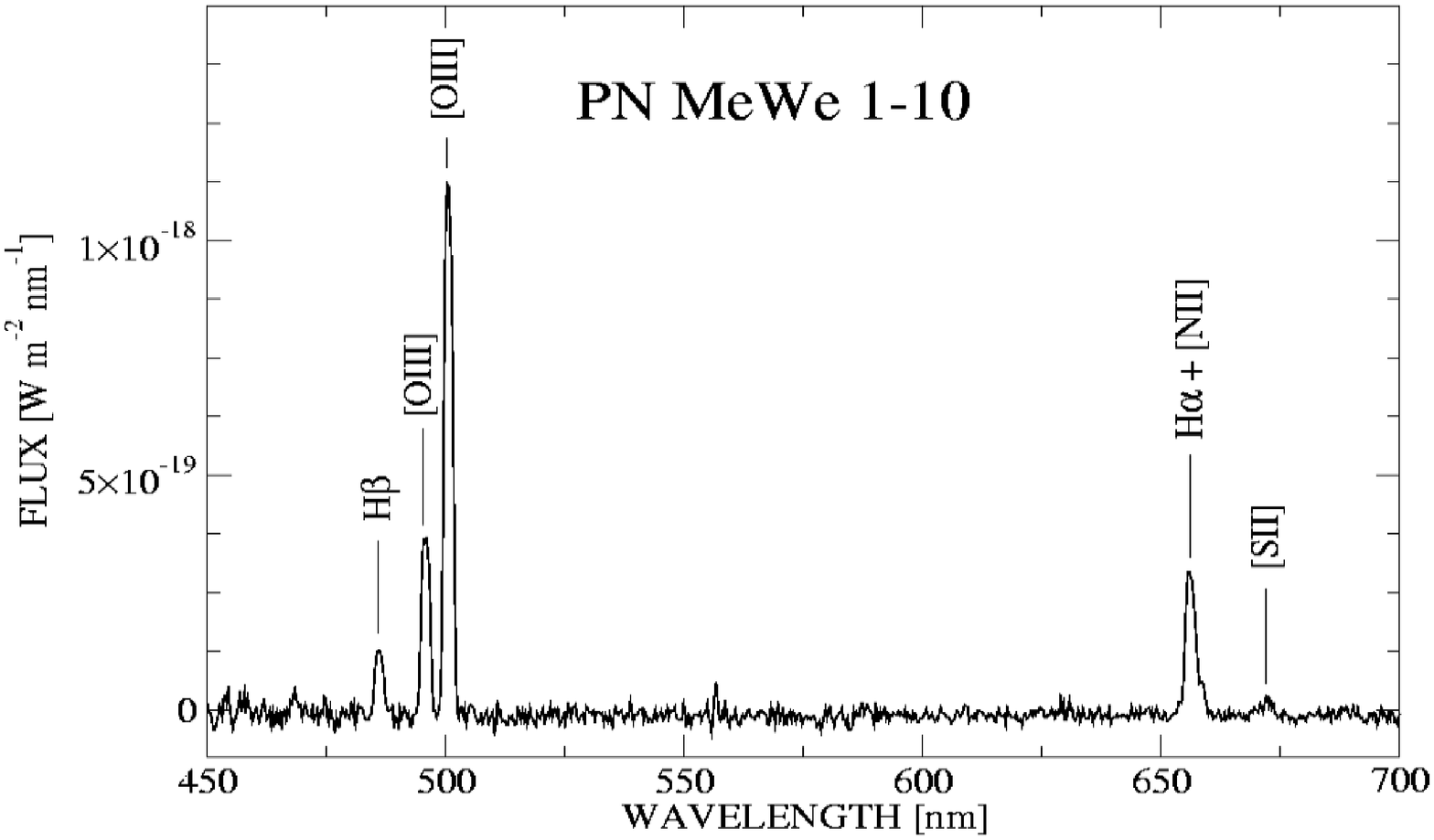}}\phantom{dslkjf}\resizebox{6.5cm}{!}{\includegraphics{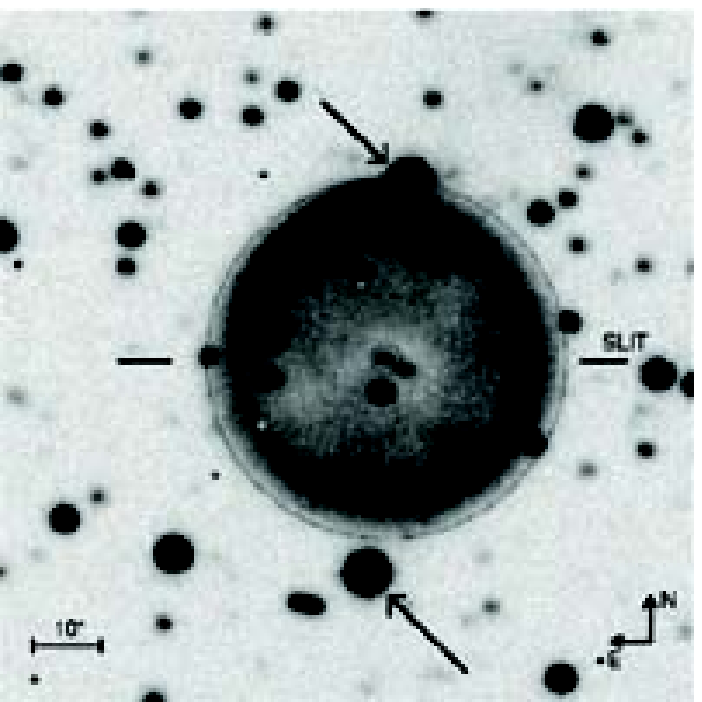}}}

\caption{Spectrum (left) and image (right) of PN MeWe 1-10. The
exposure time was 900 seconds for the spectrum as well as for the
image. The arrows in the image mark the stars used as astrometric
calibrators. The images are given in sky orientation (E is left, N
is up). The inner circle encloses the central star of the PN and
the outer circle shows the almost perfectly round shape of this
nebula. The NE enhancement might be due to ISM interaction, but as
the slit of the spectrum does not cover this region, we cannot be
sure. The offset of the outer ring with respect to the CSPN
strongly encourages this interpretation.} \label{S110}
\end{figure*}

\phantom{blabla}\relax

\begin{figure*}[ht]
\centerline{\resizebox{11.0cm}{!}{\includegraphics{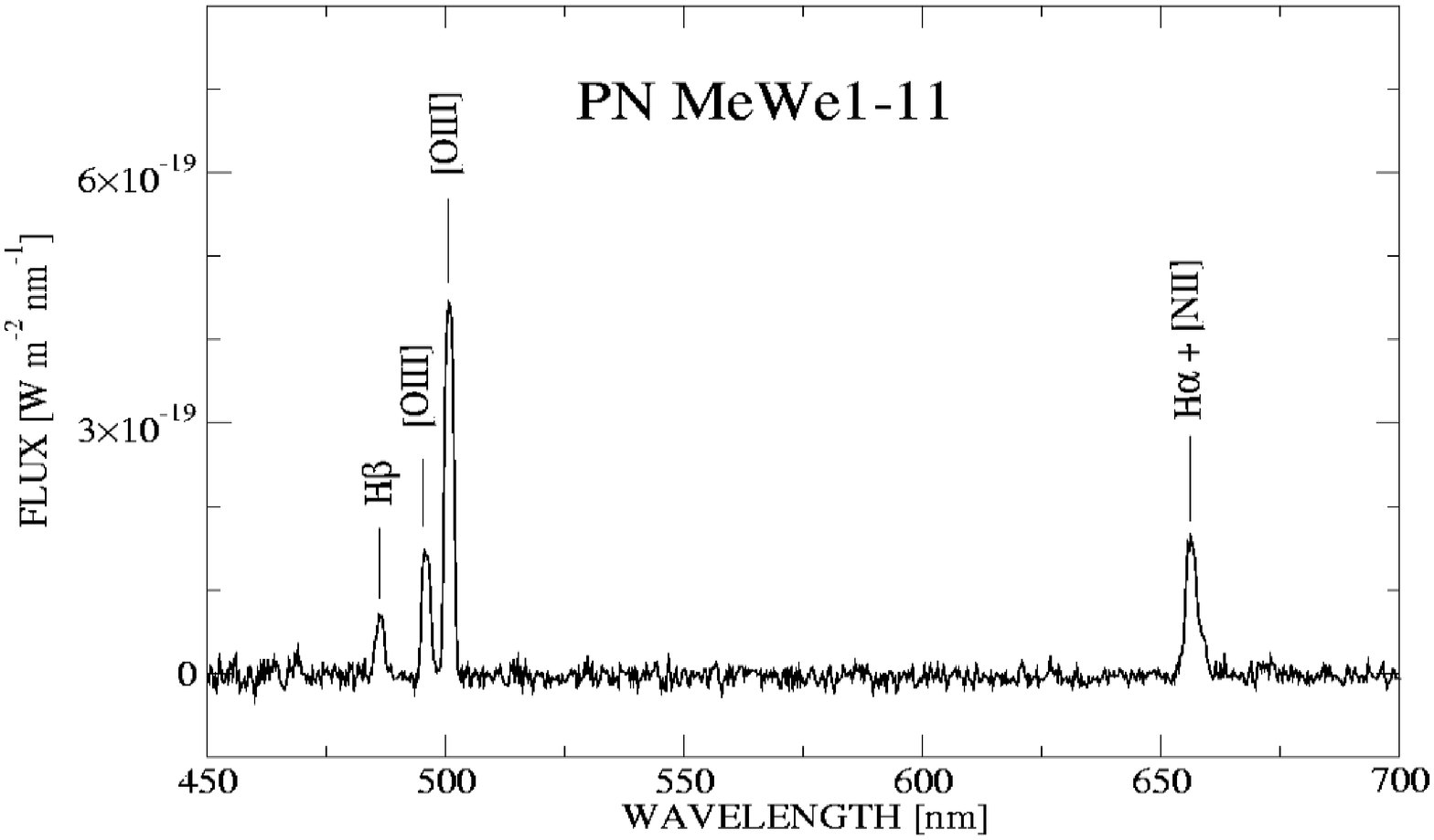}}\phantom{dsllkji}\resizebox{6.5cm}{!}{\includegraphics{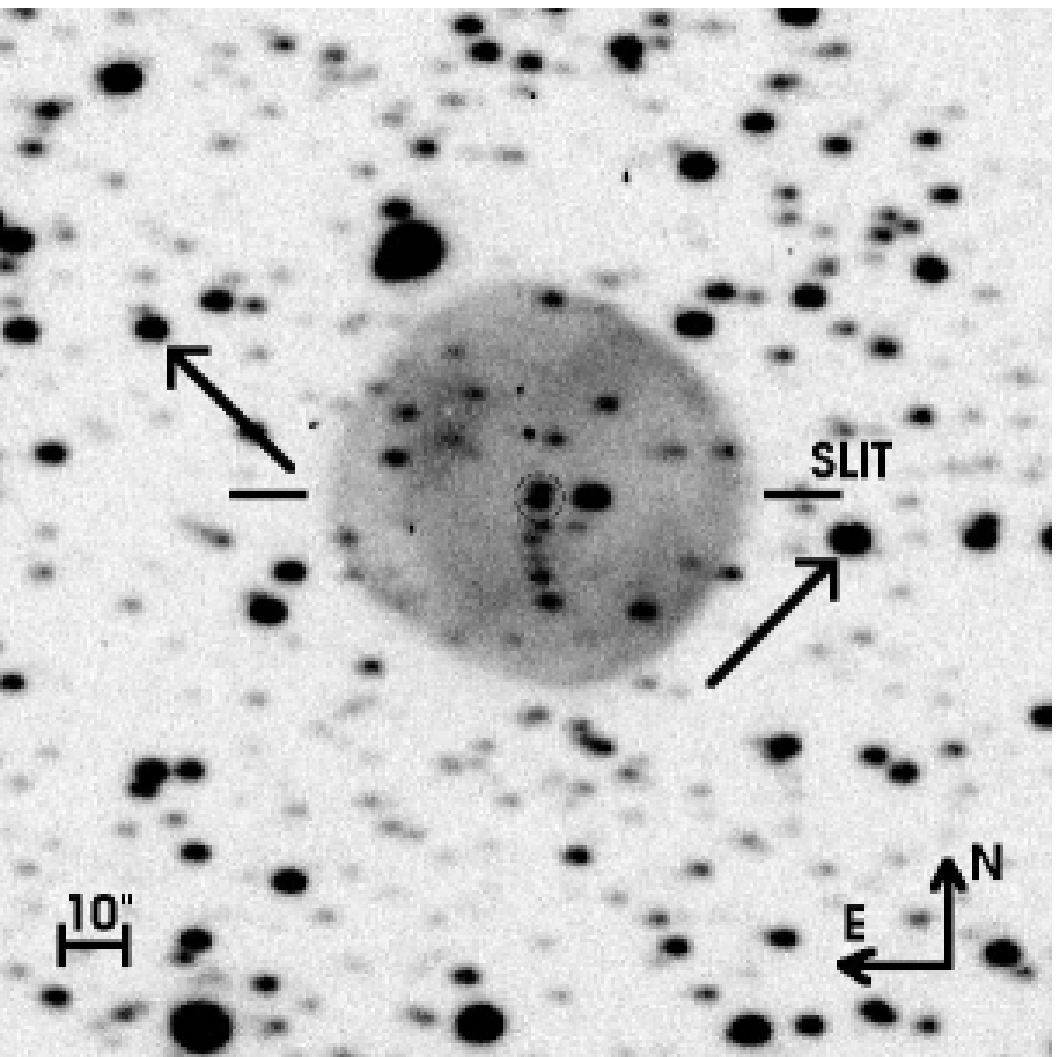}}}
\caption{Spectrum (left) and image (right) of PN MeWe 1-11. The
exposure times were 900 seconds for the spectrum and for the
image. The marks are as described in Fig.~\ref{S110}. The western
edge is clearly enhanced. This is most likely due to ISM
interaction. But including the somewhat weaker SE dip implies that
it might be a signature of a cylindrical structure (see Fig.
\ref{shock} ).} \label{S111}
\end{figure*}
\phantom{blabla}\relax

\begin{figure*}[ht]
\centerline{\resizebox{11.0cm}{!}{\includegraphics{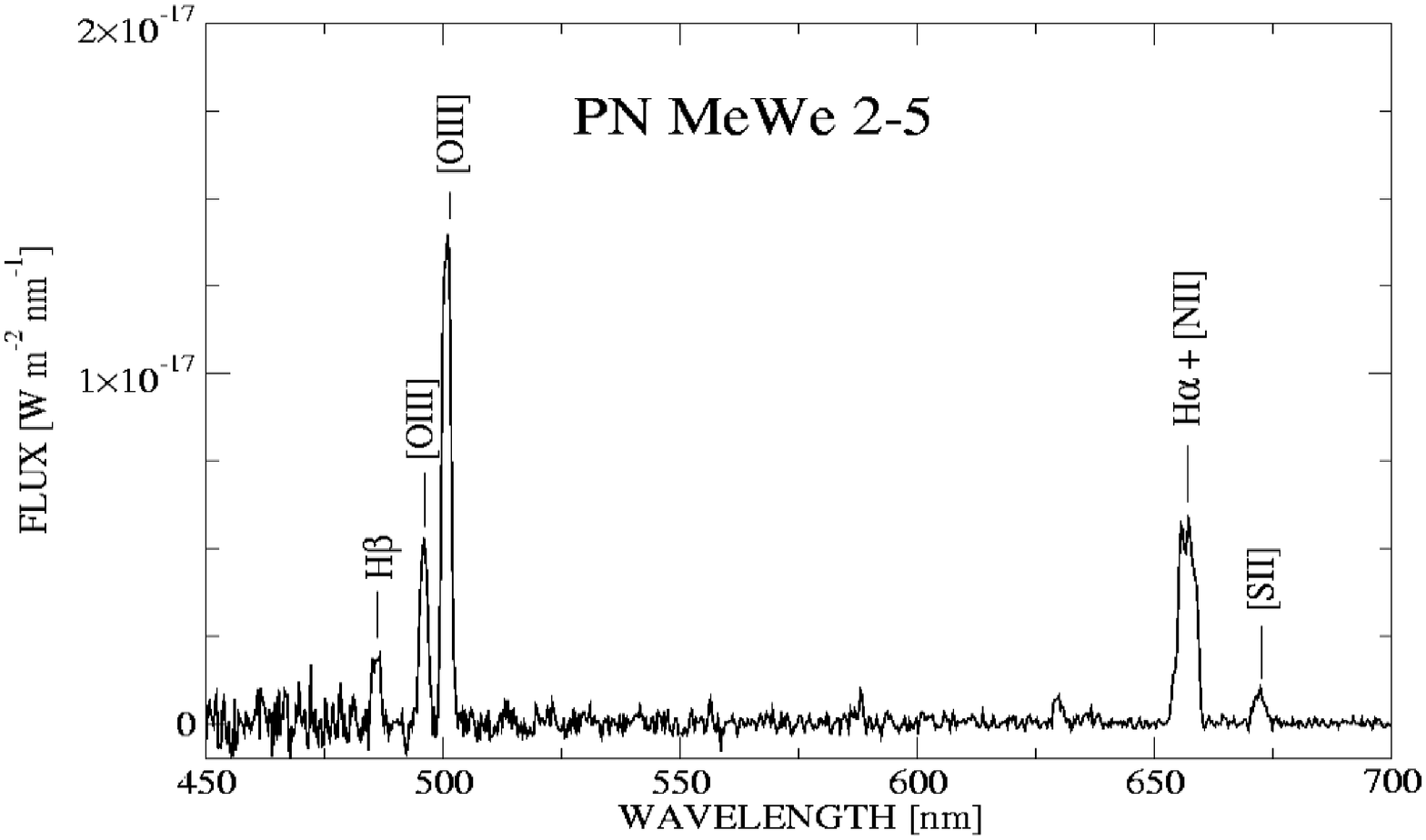}}\phantom{dslkfag}\resizebox{6.5cm}{!}{\includegraphics{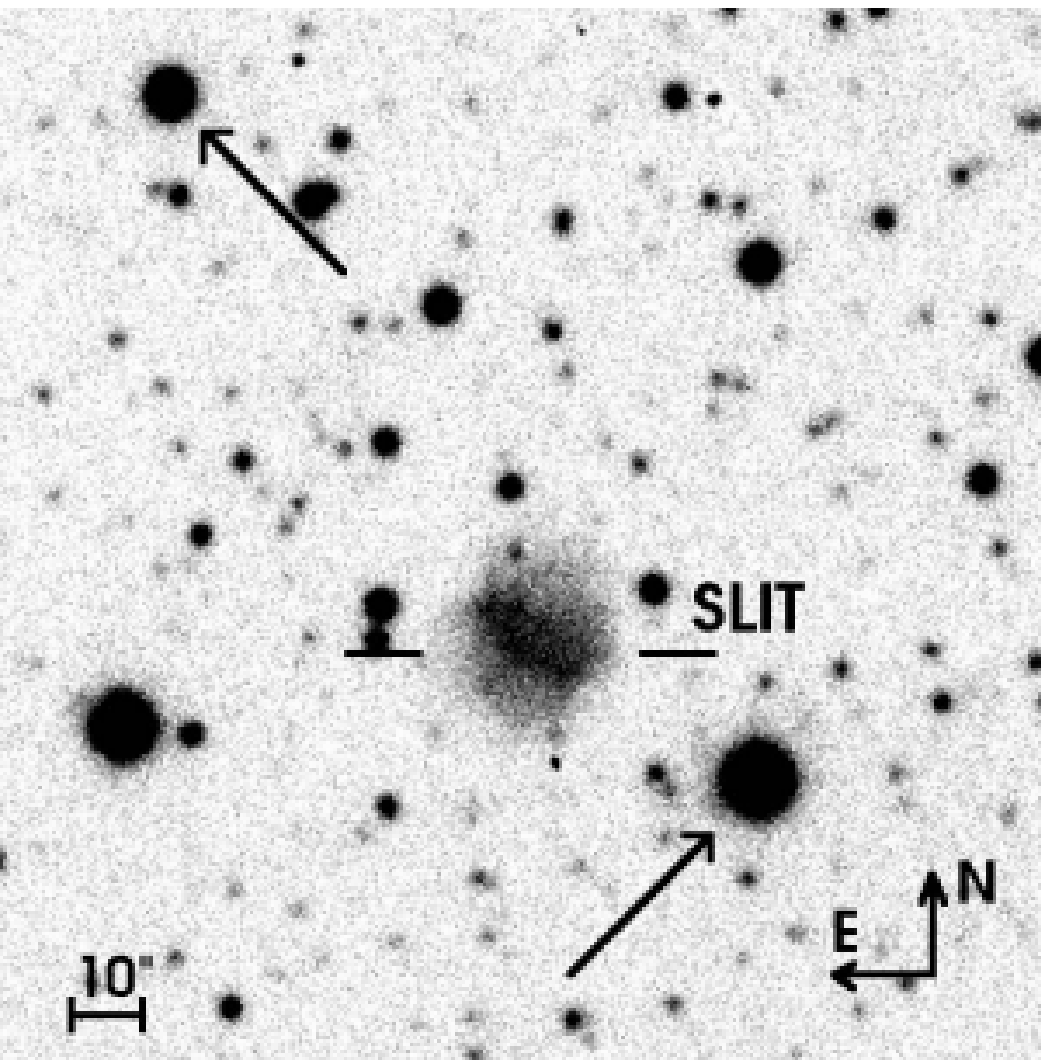}}}

\caption{Spectrum (left) and image (right) of PN MeWe 2-5. The
exposure times were 900 seconds for the spectrum and for the
image. The marks are as described in Fig.~\ref{S110}.} \label{S25}
\end{figure*}

\section{Results}

\subsection{Astrometry and cross-identification}

To obtain accurate astrometry of the PNe the narrow band
[\ion{O}{III}] image was used. In the case of \mbox{MeWe 1-10} and
\mbox{MeWe 1-11} the stars selected by Melmer \& Weinberger
(\cite{MeWe}) as candidates for the central stars of the PN
(CSPNe) were selected. Our spectra indeed show them as very hot
blue objects with very weak lines typical of CSPNe. For \mbox{MeWe
2-5} the geometric center of the central bar at the [OIII] direct
imaging frame was selected. The DFOSC has, according to the
manual, a resolution of $0\farcs39$ per pixel. We measure
$0\farcs3929$ $\pm$ $0\farcs00046$. The FWHM on the images used
for the astrometry were within the range of $2\farcs36$ to
$3\farcs54$. We used only the central part of the image with a
radius of 2$^{\prime}$ around the target. This provides us with a
highly distortion--free image. Astrometric calibrators were taken
from the USNO CCD Astrometric Catalogue (Zacharias et al.
\cite{UCAC}). This catalogue contains southern sources with an
accuracy of about 20~mas in the red magnitude range 10$^{\rm m}$
$<$ $m$ $<$ 14$^{\rm m}$ and still has an accuracy of about 70~mas
at the limit of 16$^{\rm m}$. We used the two most nearby stars
for each target to obtain the astrometry (marked in
Fig.~\ref{S110}; Fig.~\ref{S111} and Fig.~\ref{S25}). The results
are presented in Tab.~\ref{mean_tab}. According to Andersen \&
Kimeswenger (\cite{Ki02}) the error of our coordinates is assumed
to be 110 mas. Thus our coordinates are more accurate than the one
given by Kimeswenger (\cite{Ki01}) that had an rms of about
1\arcsec\ and had due to the GSC reference frame up to 2\arcsec\
of systematic effects.

\begin{table*}[ht]
\caption{Basic data for the PNe investigated} \label{mean_tab}
\begin{tabular}{lllllccc}
\hline
Name & PN G (according & GPN (according & $\alpha_{\rm J2000}$ & $\delta_{\rm J2000}$ & size &  stat. dist. & radius  \\
 & Acker et al. \cite{ESO}) & Kimeswenger \cite{Ki01}) &  &  & [\arcsec] &  [kpc] &  [pc] \\
\hline
\hline
MeWe 1-10 & {\tt PN G336.9-11.5} & {\tt GPN G336.98-11.58} & {\tt 17$^{\rm h}$34$^{\rm m}$28\fs18} & {\tt -54$^{\circ}$28\arcmin57\farcs4} & 76\arcsec & 2.9  & 0.53 \\
MeWe 1-11 & {\tt PN G345.3-10.2} & {\tt GPN G345.32-10.21} & {\tt 17$^{\rm h}$52$^{\rm m}$47\fs09} & {\tt -46$^{\circ}$42\arcmin00\farcs4} & 69\arcsec & 3.1  &  0.51\\
MeWe 2-5 & & {\tt GPN G340.93+03.75} & {\tt 16$^{\rm h}$34$^{\rm m}$49\fs60} & {\tt -42$^{\circ}$03\arcmin45\farcs1} & 28\arcsec & 8.0  & 0.54 \\
\hline
\end{tabular}
\end{table*}

\subsection{Spectroscopy}

\begin{figure}[h]
\centerline{\resizebox{8.8cm}{!}{\includegraphics{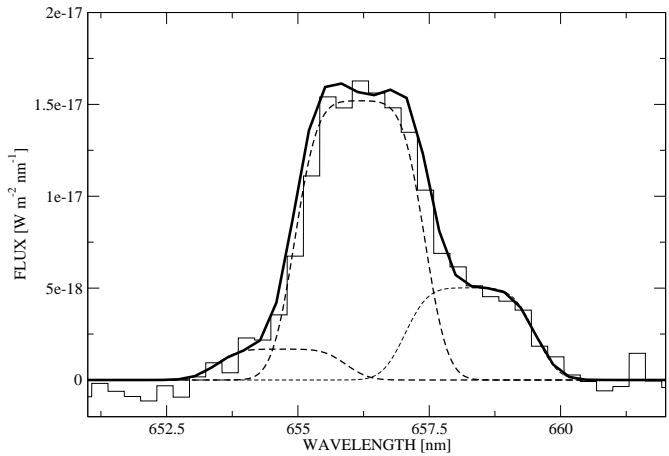}}}

\caption{The "deconvolution" of the H$\rm \alpha$ and [\ion{N}{II}] lines in
case of We 1-11. The thin line gives the data, the thin dashed lines the
individual lines of the applied fit, using the profile of a nearby
night-sky-line, using fixed positions and a fixed ratio of
[\ion{N}{II}]$_{658.4}$/[\ion{N}{II}]$_{654.8}$. The thick line gives the total
fit. } \label{linefit}
\end{figure}

All three objects are not associated with a known radio survey
source nor have IRAS counterparts been found. The relative errors
of the individual lines and the error of the line ratios were
estimated using different regions of the spectrum along the slit
and the accuracy of the standard star. As the variations lie
within a few percent, a conservative estimate gives an error of
10-15\% for the line ratios. Since the H$\rm \alpha$ and
[\ion{N}{II}] lines are not detached in our spectra, we had to
deconvolve them. We derived the line profile from the isolated
night-sky-line at 630~nm. Using this profile, the known positions
of the lines and the quantum mechanically fixed line ratio of the
[\ion{N}{II}] lines a restricted fit, only having the ratio H$\rm
\alpha$/$\sum$[\ion{N}{II}] as a free parameter, was applied. The
resulting fit is given in Fig.~\ref{linefit}. The fit quality
gives an error estimate comparable to that of the isolated lines
for H$\rm \alpha$ and [\ion{N}{II}]$_{658.4}$.

To deredden the frames we used the interstellar extinction curve
of  Savage \& Mathis~(\cite{ext}). The error estimate for the
lines results in an error in the reddening. The statistical
distance and the radius of the nebulae have been calculated from
the 5 GHz surface brightness data (Schneider \& Buckley
\cite{Dist1}). We estimated the 5 GHz data from the H$\rm \beta$
surface brightness along the slit (Cahn et al. \cite{Dist2}),
because no radio data were available.

\medskip
\noindent{\bf{MeWe 1-10}} (Fig.~\ref{S110}, Tab.~\ref{line_tab})
is a roughly round PN and it´s central star is well centered. From
the Balmer decrement we find considerable interstellar reddening
of E(B-V)~=~0\fm262, as is not unexpected for this galactic
region. The ratio of the [\ion{S}{II}] doublet, which is very
faint and also not well detached, indicates a density of $\leq
200~cm^{-3}$ clearly showing that the object is in a late stage of
evolution.

\begin{table}[h]
\caption{PN MeWe 1-10 Line identifications} \label{line_tab}
\begin{tabular}{lll}
\hline
Line [nm] & observation & model \\
& (dereddened)& \\
\hline \hline
H$\beta_{486.1}$ & 100 & 100  \\
$\rm [OIII]_{495.9}$ & 303 & 311  \\
$\rm [OIII]_{500.7}$ & 927 & 897  \\
$\rm [NII]_{654.8}$ & 18 & 18  \\
H$\rm \alpha_{656.3}$ & 287 & 287  \\
$\rm [NII]_{658.4}$ & 53 & 52  \\
$\rm [SII]_{671.7}$ & 18 & 28 \\
$\rm [SII]_{673.1}$ & 12 & -  \\
\hline
E(B-V) & 0\fm26$\pm$0\fm05 &   \\
CS temp. [K] & & 69000$\pm$3000\\
CS lum. [L$_\odot$] & & 350$\pm$100 \\
\hline
\end{tabular}
\end{table}

\begin{figure}[h]
\centerline{\resizebox{8.8cm}{!}{\includegraphics{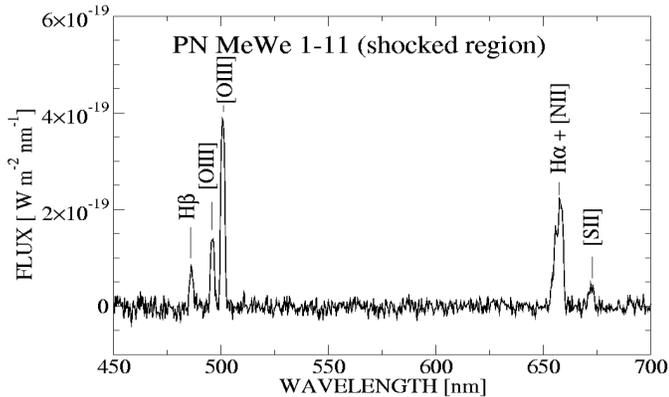}}}

\caption{The spectrum of PN MeWe 1-11 at the western edge. The
spectrum changes abruptly with respect to the other parts of the
nebula (see text).} \label{shock}
\end{figure}

\noindent{\bf{MeWe 1-11}} (Fig.~\ref{S111},  Tab.~\ref{line_tab1})
is a box-shaped PN. It shows a pronounced brightness enhancement
to the north-west in [\ion{N}{II}], that is less prominent in the
[\ion{O}{III}] image (Fig.~\ref{S111}). A possible explanation for
this peculiarity is that MeWe 1-11 is interacting with the
interstellar medium. As shown in Fig.~\ref{shock} the spectrum
changes there significantly towards partly deionized stages. While
the ionization stage, defined by [OIII] / H$_\beta$, does not
change [NII] and [SII] are enhanced abruptly. This is typical of
ISM interacting regions (Furlan \cite{Elise}; Kerber et al.
\cite{elise3}). In the diagnostic diagram of Garcia Lario et al.
(\cite{GL91}) it strongly moves towards the regions of shocked gas
(see e.g. Zanin \& Weinberger \cite{ZW}), but it still carries the
signature of a mainly photoionized PN. The Balmer decrement gives
no interstellar reddening. As the line of sight leaves the
galactic plane and as the Gould Belt clouds are above the plane in
this direction this low reddening is possible. The [\ion{S}{II}]
doublet is not detectable, but we estimate from the surface
brightness, that is comparable to that of MeWe 1-10 at a similar
distance, that the density is also $\leq 200~cm^{-3}$.

\begin{table}[h]
\caption{PN MeWe 1-11 Line identifications} \label{line_tab1}
\begin{tabular}{llll}
\hline
Line [nm] & main region & western edge & model \\
& (deredd.)& (deredd.)\\
\hline \hline
H$\beta_{486.1}$ & 100 & 100 & 100  \\
$\rm [OIII]_{495.9}$ & 221 & 205 & 239  \\
$\rm [OIII]_{500.7}$ & 690 & 591 & 689   \\
$\rm [NII]_{654.8}$ & 30 & 115 & 30   \\
H$\alpha_{656.3}$ & 269 & 222 &  287  \\
$\rm [NII]_{658.4}$ & 89 & 344 & 89 \\
$\rm [SII]_{671.7}$ & - & 51 & - \\
$\rm [SII]_{673.1}$ & - & 34 & - \\
\hline
E(B-V) & $\approx$0\fm0 & &   \\
CS temp. [K] & & & $\!\!\!\!\!\!$70000$\pm$5000 \\
CS lum. [L$_\odot$] & & & $\!\!\!\!\!\!$206$\pm$100 \\
\hline
\end{tabular}
\end{table}

\noindent{\bf{MeWe 2-5}} (Fig.~\ref{S25},Tab.~\ref{line_tab2})  is
a bipolar PN. The interstellar reddening has been found to be
E(B-V)~=~1\fm214. According to the lower galactic latitude of the
object, this  is to be expected. The ratio of the [\ion{S}{II}]
doublet, also very faint and not well detached in this object,
indicates a density of $\leq 200~cm^{-3}$. This indicates that the
object is in a late stage of evolution, like the other two
nebulae. The features at 589~nm and 630~nm seem to be relicts of
the night-sky-lines.

\begin{table}[h]
\caption{PN MeWe 2-5 Line identifications} \label{line_tab2}
\begin{tabular}{ll}
\hline
Line [nm] & observation \\
& (dereddened)\\
 \hline \hline
H$\beta_{486.1}$ & 100  \\
$\rm [OIII]_{495.9}$ & 251  \\
$\rm [OIII]_{500.7}$ & 735  \\
$\rm [NII]_{654.8}$ &  89 \\
H$\alpha_{656.3}$ & 279  \\
$\rm [NII]_{658.4}$ & 265  \\
$\rm [SII]_{671.7}$ & 42  \\
$\rm [SII]_{673.1}$ & 29  \\
\hline
E(B-V) & 1\fm21$\pm$0\fm05 \\
\hline
\end{tabular}
\end{table}

\section{Photoionisation model}

To get an idea of the temperature and the  luminosity of the
central star we modelled these nebulae with the CLOUDY code
(Ferland \cite{cloudy}). The abundances were used as in the
standard PN of CLOUDY. Only sulphur had to be lowered to the solar
value. Since the [\ion{S}{II}] doublet is very faint we based our
modelling on the relative line fluxes of [\ion{O}{III}]$_{500.7}$
and [\ion{N}{II}]$_{658.4}$. We used the NLTE central star models
of Rauch (\cite{rauch1}, \cite{rauch2}). As found by Armsdorfer et
al. (\cite{CSPN}) the use of real stellar photosphere models is
critical for the parameters of the central stars. Assuming that
the density is between 100 and 200 cm$^{-3}$ according to the
observed ratio
$\frac{[\ion{S}{II}]_{671.7}}{[\ion{S}{II}]_{673.1}}$, we used
densities of 100, 150 and 200~cm$^{-3}$. Since the radius
calculated by the method of Schneider \& Buckley (\cite{Dist1}) is
quite large, we varied also the radius from 0.5 to 1 times the
calculated radius (resp. varied the distance).
\begin{figure}[h]
\centerline{\resizebox{8.8cm}{!}{\includegraphics{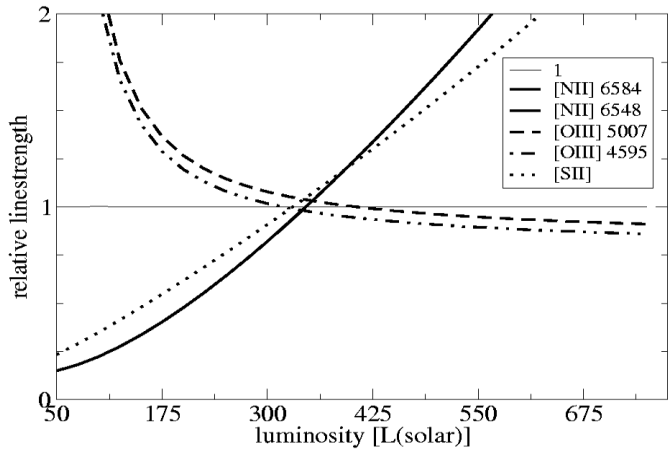}}}
\caption{The fit of a CLOUDY model (MeWe 1-10). On the abscissa is
the luminosity in units of [L$_\odot$] and on the ordinate the
ratio of ${\rm \frac{linestrenght (observed)}{linestrength
(modeled)}}$.}  \label{CLOU}
\end{figure}
Supposing a constant expansion velocity of the objects of  20 to
30 km/s, typical for such round objects (Weinberger
\cite{radvel}), we are able to calculate the age of the nebulae
from their radii. Thus it was possible to compare the luminosity
and the radii of our simulations with models of evolutionary
tracks of central stars of PNe (Bloecker \cite{blocker},
Vassiliadis \& Wood \cite{wood}). This comparison indicates that
the radii, and due to this also the distance, should be smaller by
a factor of two from the calculation above. Also the density seems
to be rather 100~cm$^{-3}$ than 200~cm$^{-3}$. In Fig.~\ref{CLOU}
the result of one of these models is shown. Because PN MeWe 2-5 is
a bipolar nebula, CLOUDY is not a proper tool to model this
object. We thus decided not to model it.

\section{Conclusion}

We have spectroscopically confirmed the nature of three PNe,
especially of the possible PN MeWe 2-5. Even this small sample
already shows that each PN is unique. In particular the old,
extended PNe found in the optical surveys tend to be intresting
due to their late evolutionary stage; for an ever--increasing
number of these PNe, signs of an interaction with the ISM are
being discovered.

The photoionisation models suggest a significantly smaller
distance to the objects than those found using statistical
distances.

The nebulae MeWe 1-10 and MeWe 1-11 lie below the domain of the
tracks  from Bloecker (\cite{blocker}) and Vassiliadis \& Wood
(\cite{wood}). This results in low mass CSPN and thus suggests low
mass progenitors. Our model does not lead us to a low abundance of
at least CNO elements. This thus does not support the hypothesis
of Soker (\cite{soker}) that round PNe may be formed only from low
abundance progenitors.

Further detailed high-resolution spectroscopic investigations of
these central stars, so as to be able to model them as in
Napiwotzki (\cite{napi1} \& \cite{napi2}) are strongly encouraged
to fix the CSPN parameters. This then will allow us to study the
nebula in more detail by fixing parameters like $T_{\rm CSPN}$ and
distance.

\begin{acknowledgements}
We thank the anonymous referee for her/his careful reading of the
original manuscript and for useful comments.
\end{acknowledgements}

\end{document}